\documentclass{article}
\usepackage{PRIMEarxiv}
\usepackage[utf8]{inputenc} 
\usepackage[T1]{fontenc}    
\usepackage{hyperref}       
\usepackage{url}            
\usepackage{booktabs}       
\usepackage{amsfonts}       
\usepackage{nicefrac}       
\usepackage{microtype}      
\usepackage{lipsum}
\usepackage{fancyhdr}       
\usepackage{graphicx}       
\graphicspath{{media/}}     
\usepackage{tabularx}
\usepackage{multirow}
\title{A multi-scale assessment for managing coastal geomorphic changes in southwestern Lake Michigan
\thanks{\textit{\underline{Citation}}: 
\textbf{Lu, B., Wang, W., Jordan, N., Wright, D., Bechle, A., Zoet, L., \& Wu, C. (2025). A multi-scale assessment for managing coastal geomorphic changes in southwestern Lake Michigan. Journal of Environmental Management, 395, 127878. https://doi.org/10.1016/j.jenvman.2025.127878}} 
}

\author{
Boyuan Lu \\
Department of Civil and Environmental Engineering \\
University of Wisconsin, Madison \\
Madison\\
  \texttt{blu38@wisc.edu} \\
   \And
   Wei Wang \\
   Department of Civil and Environmental Engineering \\
University of Wisconsin, Madison \\
Madison\\
  \texttt{wwang487@wisc.edu} \\
   \And
   Nick Jordan \\
   Department of Civil and Environmental Engineering \\
University of Wisconsin, Madison \\
Madison\\
  \texttt{nbjordan7@gmail.com} \\
   \And
   Daniel Wright \\
   Department of Civil and Environmental Engineering \\
University of Wisconsin, Madison \\
Madison\\
  \texttt{danielb.wright@wisc.edu} \\
   \And
   Adam Bechle \\
   Wisconsin Sea Grant \\
University of Wisconsin, Madison \\
Madison\\
  \texttt{bechle@aqua.wisc.edu} \\
  \And
   Lucas Zoet \\
   Department of Geoscience \\
University of Wisconsin, Madison \\
Madison\\
  \texttt{lzoet@wisc.edu} \\
  \And
Chin Wu \\
   Department of Civil and Environmental Engineering \\
University of Wisconsin, Madison \\
Madison\\
  \texttt{chin.wu@wisc.edu} \\
}

\begin{document}
\maketitle

\begin{abstract}
Understanding coastal geomorphic change is essential for advancing the United Nations Sustainable Development Goals (SDGs) through a multi-scale coastal management framework. In particular, characterization of coastal geomorphic change across multiple spatial and temporal scales can provide essential insights and context-specific knowledge that can inform and empower local communities. In this study, we present a multi-scale assessment of coastal geomorphic change in southwestern Lake Michigan in the Laurentian Great Lakes. Three spatial scales—county, reach, and transect—and two temporal scales—long-term and short-term—were examined using nine sets of historical aerial imagery spanning 1937 to 2020. The site-averaged long-term (1937-2020) change rates for the bluff crest, bluff toe, and shoreline were –0.22, –0.17, and –0.16 m/year, respectively. In the short term (1995-2020), the corresponding rates were –0.22, –0.15, and –0.32 m/year, indicating an increasing shoreline erosion in recent years. The coastal geomorphic changes at county, reach, and transect scales were further characterized, showing regional and localized distributions of coastal erosion in our study sites. The mechanisms driving coastal change—particularly wave impacts—were also examined to assess their correlation with coastal geomorphic change across different spatial scales. The results indicate that wave impacts influence coastal environments at certain scales more strongly than at others. Several erosion “hotspots” were examined to identify local factors contributing to severe site-specific erosion. Lastly, the spatial uniformity of coastal geomorphology was examined between the county and reach scales. Overall, the findings suggest that multi-scale analyses provide a valuable insight for effective management of coastal geomorphology.
\end{abstract}


\section{Introduction}
Coastal geomorphic change (CGC) describes how coastal areas evolve through shoreline, beach, and bluff accretion or erosion, which significantly affects property values and public safety \cite{vitousek2024scalable,allen2019linking}. Globally, over 2.8 billion people live within 100 km of coasts, making them vulnerable to coastal hazards \cite{martinez2007coasts,cosby_accelerating_2024}.It is reported that, based on satellite imagery from 1984 to 2016, approximately 24\% of the world’s sandy beaches experienced erosion \cite{luijendijk_state_2018}, while 28\% underwent accretion, highlighting the global scale of CGC. In the Laurentian Great Lakes, CGC affects roughly 60\% of the shoreline and 30 million residents \cite{mickelson1977shoreline,mickelson2004erosion,brown_factors_2005,jackson_coastal_2013}. Recent studies have found that coastal geomorphic changes in the Great Lakes region have increased, due primarily to water level increases of roughly 2 meters since 2014 \cite{gronewold_hydrological_2016,gronewold_tug--war_2021}. The rapid increment of water level led to accelerated shoreline erosion and bluff collapse, causing damages to lakeside properties . Meanwhile, coastal habitat loss was also found due to the CGC caused by rapid rising water level \cite{theuerkauf_rapid_2021}. Given the increasing risks of CGC caused by rising water level, better assessment of coastal geomorphic changes is essential for effective management and hazard reduction in the Great Lakes \cite{lawrence1994natural}.

CGC results from a complex interplay of factors operating across diverse spatial and temporal scales \cite{lollino2021multi}. Over long-term periods (usually over years), CGC is shaped by processes such as lake-level fluctuations \cite{meadows_relationship_1997,brown_factors_2005,swenson_bluff_2006}, freeze–thaw cycles \cite{vallejo1981stability,bernatchez2011development,roland2021seasonality}, ice forces \cite{barnes1994influence,bamasoud2012impact,volpano2025modeling}, and lakebed downcutting \cite{davidson2000effects,fuller2002bank}. In contrast, short-term variations—lasting from hours to days—can arise from shallow or deep-seated slope failures \cite{edil_mechanics_1980,quinn2010identifying} and groundwater seepage \cite{collins_processes_2008, brooks2012deriving}. Spatially, large-scale processes such as littoral sediment transport regulate sediment budgets across hundreds of meters to kilometers \cite{harley2011reevaluation}. Localized changes occur at smaller scales, often between meters and hundreds of meters. For instance, short-lived storms can intensify wave attack, rapidly eroding beaches locally \cite{ruggiero_wave_2001,swenson_bluff_2006}, whereas human activities such as shoreline armoring can deplete nearshore sediment reservoirs and cause flanking erosion down-drift \cite{lin_field_2014}. Additionally, local factors including bluff geology \cite{swenson_bluff_2006} and land-cover alterations associated with development \cite{borzi2025impact} can accelerate geomorphic change. Altogether, the interactions among these temporally and spatially variable processes underscore the importance of multi-scale analyses for understanding and managing CGC \cite{ells2012long}.

Multiscale analysis is a widely adopted approach in coastal regions, offering critical insights for effective shoreline management and coastal resilience planning. For example, studies employed both decadal and event-scale shoreline data to identify localized erosion and accretion patterns that would otherwise be obscured at broader temporal resolutions \cite{zoysa2023analysis,marrero2024using}. Similarly, Vitousek\cite{vitousek2024scalable} demonstrated that models integrating long-term shoreline trends with wave and sea level datasets across large spatial extents significantly enhance predictability of regional coastal change. Lollino \cite{lollino2021multi} further emphasized the importance of multi-scale methodologies in capturing coastal instabilities and geomorphic thresholds. In the Great Lakes region, multiscale studies also have been conducted at various spatial and temporal resolutions, ranging from broad-scale analyses covering hundreds of kilometers of coastline \cite{lawrence1994natural,hapke_geomorphic_2013,roland2025lidar}, to more localized regional investigations \cite{mickelson1977shoreline,birkemeier1980effect,lin_field_2014,terpstra1992geometric}. Research on Great Lakes shoreline erosion spans a wide range of temporal scales, from short-term event-based or seasonal assessments to multi-decadal analyses. Short-term studies, such as Zoet \cite{zoet_analysis_2017} and Volpano \cite{volpano_three-dimensional_2020}, focus on event-scale bluff failures and seasonal to interannual variability, providing insights into rapid geomorphic responses to water-level fluctuations. Intermediate-term studies spanning several years, including Jibson \cite{jibson1994rates} and Krueger \cite{krueger2020coastal}, examine bluff evolution under variable lake levels and storm activity. Long-term, decadal-scale investigations, such as Mickelson \cite{mickelson1977shoreline}, Brown \cite{brooks2012deriving}, and Swenson \cite{swenson_bluff_2006}, quantify sustained erosion trends and identify the dominant drivers—such as wave climate, water levels, and sediment composition—over periods exceeding 30 years. 

The objective of this study is to characterize CGC across multiple temporal and spatial scales along a 125 km stretch of the southwestern Lake Michigan shoreline from 1937 to 2020. The analysis focuses on three key coastal features—the bluff crest, bluff toe, and shoreline—measured at 10 m intervals across the study area, with results aggregated into three distinct spatial scales: the state–county scale (10–50 km), the reach–region scale (1–10 km), and the transect scale (100 m). The study also adopted two temporal scales: a short-term period (1995–2020), representing recent CGC, and a long-term period (1937–2020), reflecting historical erosion over an extended timeframe. This dual-scale approach provides valuable insights for both coastal management and research purposes. Additionally, hotspot and clustering analyses are employed to assess the practical application of multi-scale approaches in coastal management. This multi-scale framework provides comprehensive insights into the spatial and temporal variability of coastal geomorphological change and its driving forces, offering a deeper understanding of the transformations occurring in the Great Lakes region. 

\section{Study site}

\begin{figure}
    \centering
    \includegraphics[width=0.8\linewidth]{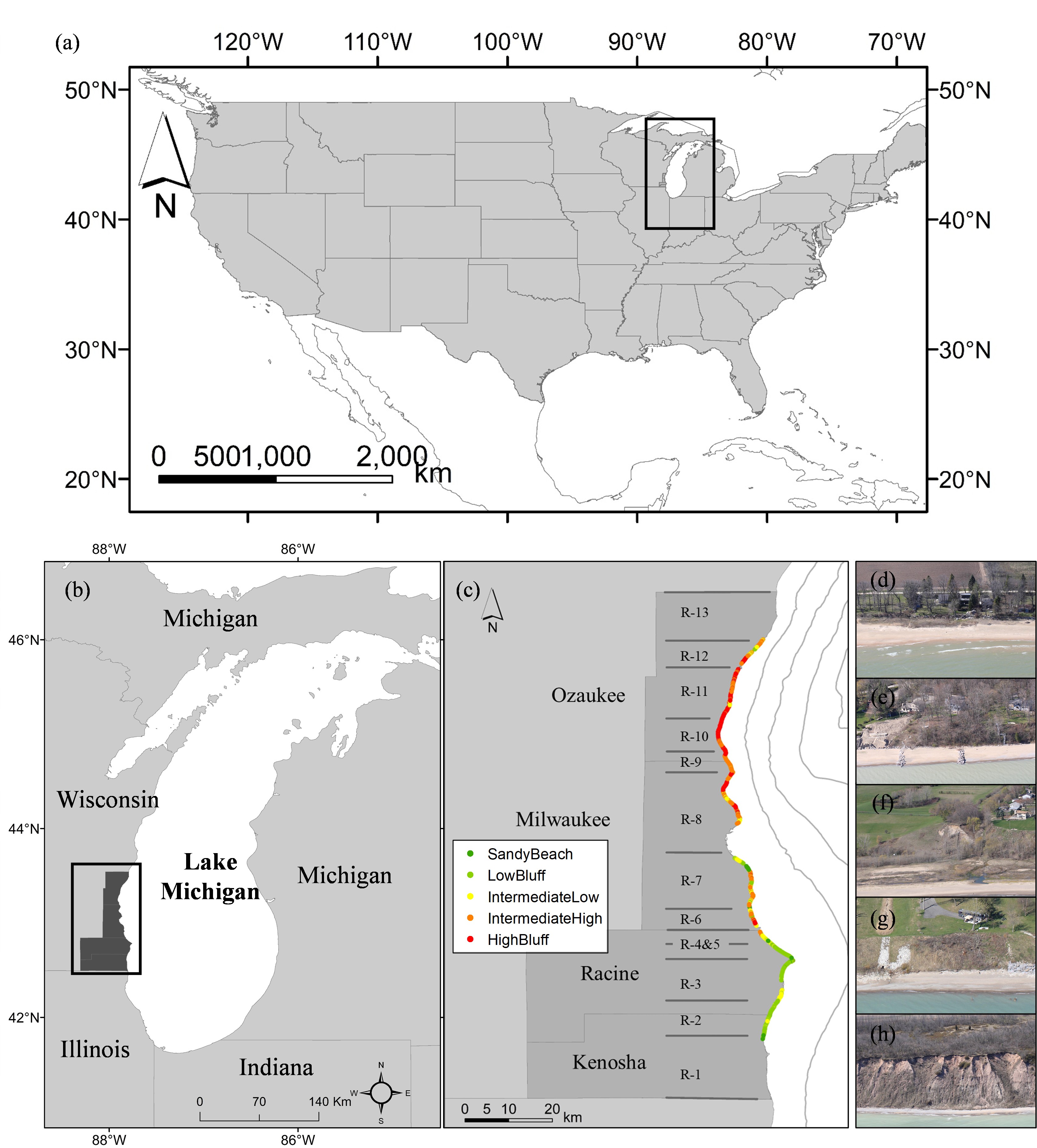}
    \caption{Study sites. (a) the location of study site in USA. (b) The
southeastern region of Wisconsin on the western coastline of Lake Michigan. (c)
Regional map of four counties and 13 reaches, denoted as R1, R2, ..., R13. (d) -
(h) are the oblique photos for different geology in study sites:  sandy shore
without bluffs, low bluff (height<10 m), intermediate low bluff (height between
10 and 20 m), intermediate high bluff (height between 20 and 30 m), and high
bluff (height>30 m). Oblique images are accessed from the Wisconsin Shoreline
Inventory and Oblique Photo Viewer
(\url{http://floodatlas.org/asfpm/oblique_viewer/})} 
\label{fig:fig2.1}
\end{figure}

The study area is located along the southwestern shore of Lake Michigan in the Laurentian Great Lakes region of the United States (Fig. \ref{fig:fig2.1}a-b). The area covers a 125-kilometer-long (77-mile-long) coastline stretching north from the Wisconsin/Illinois state boundary through Kenosha County to the northern border of Ozaukee County (Fig \ref{fig:fig2.1}c). Since the late 19th and early 20th centuries, tremendous coastal development in the form of residential properties, industries, and public infrastructure have been made in this area. We further divided the study region into thirteen reaches (Fig. \ref{fig:fig2.1}c) based on the reach classification in \cite{mickelson1977shoreline}(see geological-geomorphic setting in Table \ref{tab:tab2.1}). The geological-geomorphic setting in southeastern Wisconsin’s coastline is primarily composed of unconsolidated glacial sediments deposited during the late Wisconsin Glaciation between 25,000 and 10,000 years ago \cite{mickelson2004erosion}. Most southeastern Wisconsin’s coast consists of glacial till bluffs composed of clay, sand, and silt, stratified lacustrine sediments, and sand and gravel lenses. Several small (<100 m long) Paleozoic dolomite outcrops are present along the shore \cite{mickelson1977shoreline,mickelson2004erosion,mickelson2007wisconsin}. Fig. \ref{fig:fig2.1}b depicts the categorization of the shoreline in southeastern Wisconsin based on presence and height of bluffs. In undeveloped areas of southern Kenosha County from the Illinois border to southern end of Kenosha Harbor, a large area of beach ridge and swale zone is present (Fig. \ref{fig:fig2.1}d). Between Kenosha Harbor and Wind Point in Racine, glacial till bluffs increase to heights of 5-10 meters and are principally composed of New Berlin and Oak Creek Tills (Fig. \ref{fig:fig2.1}e). Bluffs maintain heights of 10-20 m through much of Milwaukee County (Fig. \ref{fig:fig2.1}f), except for the low-lying developed area around Milwaukee Harbor, and reach heights of 30 to 45 m in southern Ozaukee County (Fig. \ref{fig:fig2.1}h). Bluff heights decrease from 30 to 20 m north of Port Washington Harbor in Ozaukee County (Fig. \ref{fig:fig2.1}g), and a low-lying sandy shoreline stretches for 12 km north to the county border \cite{mickelson1977shoreline}.
\begin{table}[htbp]
  \caption{Definitions of coastal reaches.}
  \centering
  \setlength{\tabcolsep}{8pt}
  \renewcommand{\arraystretch}{1.15}
  \begin{tabularx}{\textwidth}{c X}
    \toprule
    \textbf{Reach ID} & \textbf{Description} \\
    \midrule
    1  & Flat, sandy, no bluff, heavily developed \\
    2  & Low bluff, heavily developed \\
    3  & Flat coast/very low bluff, sandy, mostly defended, medium development, SE orientation \\
    4  & Low bluff, heavily defended, sandy sediments, NE orientation \\
    5  & High cohesive bluff, little to no shore protection \\
    6  & High cohesive bluff, industrial land uses, increasingly defended since 1937 \\
    7  & High cohesive bluff with few structures, primarily recreational land use \\
    8  & High bluff, heavily developed residential land use \\
    9  & Donges Bay, high bluff, moderately defended, moderate developed land use \\
    10 & High bluff, moderate bluff top development and shore defense \\
    11 & High bluff, sparse bluff top development \\
    12 & High bluff, sparse bluff top development \\
    13 & Low dune/flat coast, no bluff \\
    \bottomrule
  \end{tabularx}
  \label{tab:tab2.1}
\end{table}

\section{Methods}

\subsection{Digitization of coastal geomorphology}
\begin{figure}
    \centering
    \includegraphics[width=0.8\linewidth]{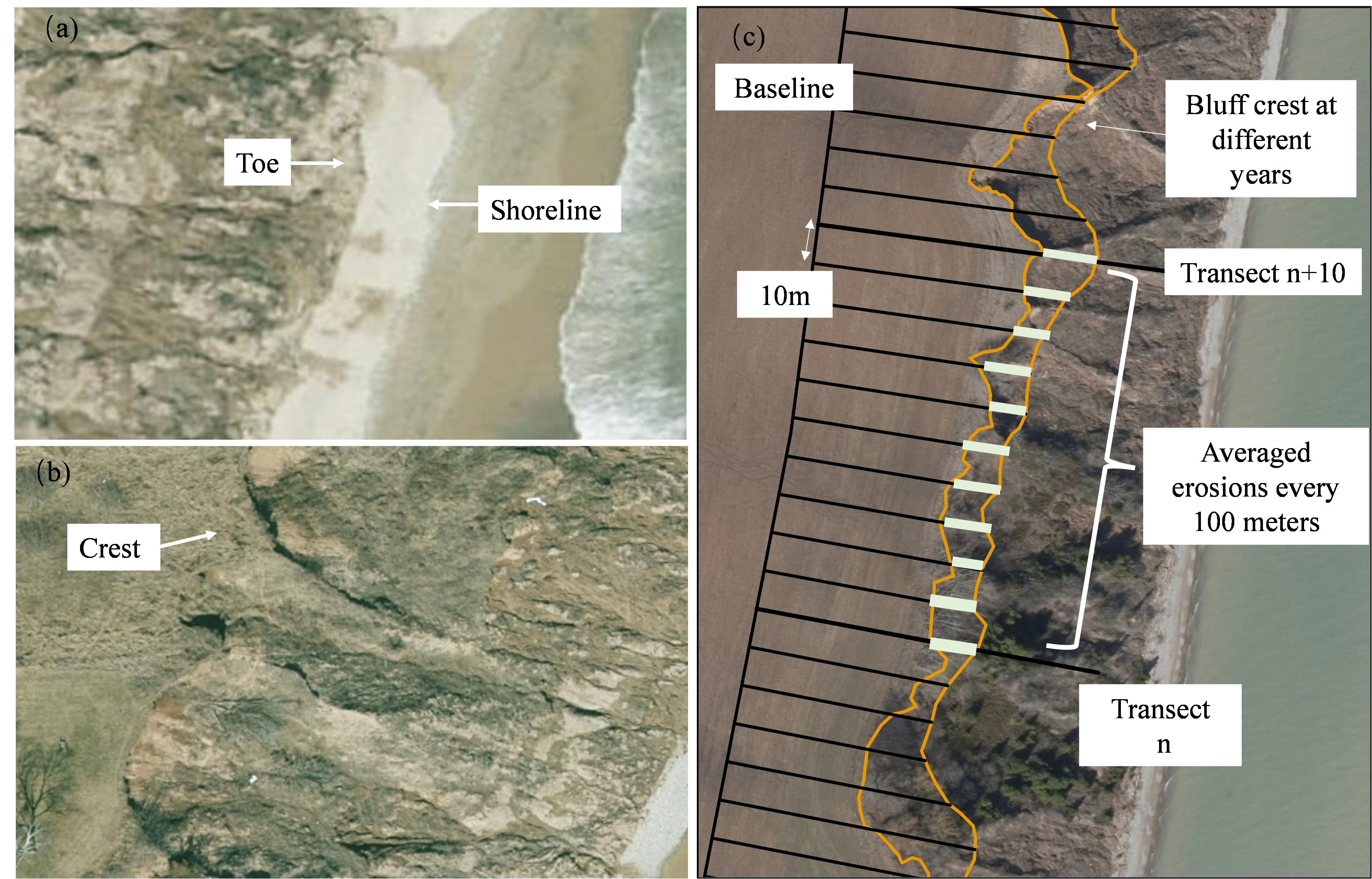}
\caption{Schematic of geomorphic feature digitization and bluff crest recession
rate calculation along 10 meter transects.} 
\label{fig:fig2.2} 
\end{figure}

A critical component of this study is the consistent digitization of key coastal features across years and alongshore positions in each photograph, focusing on shoreline, bluff toe, and bluff crest as established metrics for bluff and beach change (Zuzek et al., 2003; Brown et al., 2005; Swanson et al., 2006; Wang et al., 2025), while also incorporating beach width and bluff face slope (Fig. \ref{fig:fig2.2}). Shoreline position (Fig. \ref{fig:fig2.2}a), delineated by the High Water Line (HWL) visible as the wet-dry sand boundary in aerial imagery \cite{del_rio_error_2013}, was selected as a proxy because it can be readily and consistently identified in photographs. The bluff toe (Fig. \ref{fig:fig2.2}a) is defined as the slope break between bluff face and sandy beach, often marked by sediment or vegetation changes, while the bluff crest (Fig. \ref{fig:fig2.2}b) marks the break between bluff top and bluff face; beach width (Fig. \ref{fig:fig2.2}a) represents the horizontal distance between shoreline and bluff toe, and bluff face slope (Fig. \ref{fig:fig2.2}c) is the elevation gradient between bluff crest and toe. In sandy low-relief areas, such as southeast of Kenosha Harbor (Reach 1) and northeast of Port Washington Harbor (Reach 13), only shoreline was digitized, and no features were mapped inside harbor structures; similarly, bluff crests obscured by vegetation or development were excluded, and in locations with shore-parallel structures at the bluff toe, the toe was defined as the lakeward slope break, coinciding with shoreline when located at the waterline. Feature digitization was conducted in GIS (ArcMap) from eight sets of aerial photographs (1937–2020; Table \ref{tab:tab2.2}), selected for spatial coverage, resolution, and seasonal suitability (snow-free, leaf-off), with intervals of ~20 years between 1937–1995 and ~5 years between 1995–2020; photos were typically synoptic across the study area except for 1969–1975, which varied by county. 
To minimize errors arising from manual digitization and image georeferencing, a comprehensive quality control procedure was implemented throughout the entire process. Four annotators were involved in tasks such as image georeferencing and feature digitization, under the supervision of an expert with extensive knowledge of the local conditions at the study sites. Through this quality control approach, georeferencing errors were reduced to within 1 meter across more than 16 control points per image, while measurement uncertainties were constrained to within two pixels.
\begin{table}[ht]
  \caption{Aerial photo information for photos used in the analysis.}
  \centering
  \footnotesize
  \setlength{\tabcolsep}{4pt} 
  \renewcommand{\arraystretch}{1.2} 
  \begin{tabularx}{\textwidth}{l l l X p{2cm}}
    \toprule
    \textbf{Date} & \textbf{Photo Source} & \textbf{Spatial Extent} & \textbf{Photo Description} & \textbf{Photo Scale/Resolution (m)} \\
    \midrule
    August 1937   & USDA            & Entire Study Area & B\&W Aerial Photo       & 1--1.6 \\
    May--July 1956 & USDA           & Entire Study Area & B\&W Aerial Photo       & 1 \\
    June 1969     & USDA            & Racine Co.        & B\&W Aerial Photo       & 1 \\
    August 1971   & USDA            & Ozaukee Co.       & B\&W Aerial Photo       & 1.8 \\
    April 1975    & Milwaukee Cty LIO & Milwaukee Co.   & B\&W Aerial Photo       & 0.2 \\
    March 1976    & USGS            & Kenosha Co.       & B\&W Aerial Orthophoto  & 3 \\
    April 1995    & SEWRPC          & Entire Study Area & B\&W Aerial Orthophoto  & 0.61 \\
    April 2000    & SEWRPC          & Entire Study Area & B\&W Aerial Orthophoto  & 0.31 \\
    April 2005    & SEWRPC          & Entire Study Area & Color Aerial Orthophoto & 0.31 \\
    April 2010    & SEWRPC          & Entire Study Area & Color Aerial Orthophoto & 0.31 \\
    April 2015    & SEWRPC          & Entire Study Area & Color Aerial Orthophoto & 0.15 \\
    April 2020    & SEWRPC          & Entire Study Area & Color Aerial Orthophoto & 0.15 \\
    \bottomrule
  \end{tabularx}
  \label{tab:tab2.2}
\end{table}

\subsection{Recession rate calculation }
Historical geomorphic feature positions and rates of change were measured using the Digital Shoreline Analysis System v4 (DSAS, \cite{thieler2009digital}) along transects at 10-meter intervals from a baseline running parallel to the coastline (Fig \ref{fig:fig2.2}c). An alongshore smoothing distance of 100 m reduced transect overlap and spreading at concave and convex portions of the shoreline, respectively. To minimize local errors resulting from overlapping or misaligned transects, recession rates were averaged across every 10 transects, corresponding to an interval of approximately 100 meters. Two different temporal scales were selected to calculate the recession rate: long-term (1937-2020) and short-term (1995-2020). The long-term recession rate captures historical CGC under a relative stable water level fluctuation. This also reflects the most comprehensive record derived from the full range of available imagery. Oppositely, the short-term recession rate represents decadal-scale shoreline retreat influenced by rapid water level rising in recent years. Both recession rates were calculated using linear regression rate (LRR), which accounts for the position of a given feature in every photo within the period of interest. In general, LRR is preferred when determining average rates of erosion because influences of extreme water levels or errors in any one photo set are limited \cite{thieler2009digital}. Shoreline positions were not adjusted for water level fluctuations for two primary reasons. First, the absence of historical bathymetric data at the study sites hinders accurate calibration for water level variations, particularly in estimating long-term shoreline recession rates. Second, the influence of water level fluctuations can be considered negligible at decadal scales \cite{hanrahan_attribution_2014}, as these fluctuations tend to follow interdecadal periodic patterns that average out over time.
We considered two time periods over which changes at the bluff crest, bluff toe, shoreline, and beach width are assessed: long-term (1937-2020), the longest period of record available which contains periods of high and low water levels; and short-term (1995-2020), a period which consisted of high-water levels from 1995-1998, sustained low water levels from 1999-2013, and a rapid rise in water level since 2014. Bluff recession measurements were not calculated for Reaches 1 and 13, as these areas are characterized by relatively flat, sandy backshore regions. Erosion rates are presented as negative values, which accretional or lakeward changes are presented as positive values. 

\section{Results}

\subsection{Domain and County scale CGC}
Domain- and county-scale CGC results are summarized in Table \ref{tab:tab2.3}. Over the entire study area, the average bluff crest recession rate for the 1937-2020 period is -0.220 m/year, the average bluff toe recession rate is -0.170 m/year, and shoreline recession rate is -0.16 m/year. For the 1995-2020 period, the average recession rates for the bluff crest, bluff toe, and shoreline are -0.22, -0.15, and -0.32 m/year, respectively. Among shoreline, bluff toe and bluff crest, shoreline recession rates exhibit greater variance, between -0.06 m/year in Milwaukee County to -0.40 m/year in Kenosha County from 1937-2020. With the exception of Kenosha County, county-level average shoreline recession rates from 1995 to 2020 were higher than those observed over the 1937 to 2020 period. This suggests an acceleration in shoreline retreat in recent decades. The bluff crest, bluff toe, and shoreline in all counties exhibited consistent erosional trends over both the long-term period (1937–2020) and the more recent short-term period (1995–2020). Besides the CGC, the long-term and short-term average of beach width and bluff slopes are also reported in Table \ref{tab:tab2.4} in the appendix section in supplementary document. In contrast to coastal CGC, the differences between long-term and short-term trends in beach width and bluff face slope are minimal, suggesting that these components of bluff and beach have remained relatively stable over time. 
\begin{table}[ht]
  \caption{County-averaged bluff crest, bluff toe, and shoreline recession rates. Negative values indicate bluff erosion, while positive values indicate bluff advance.}
  \centering
  \small
  \setlength{\tabcolsep}{4pt}
  \renewcommand{\arraystretch}{1.2}
  \begin{tabularx}{\textwidth}{l *{3}{>{\centering\arraybackslash}p{2.2cm}} *{3}{>{\centering\arraybackslash}p{2.2cm}}}
    \toprule
    \multicolumn{1}{c}{\textbf{County}} &
    \multicolumn{3}{c}{\textbf{1937--2020 Recession Rate (m/yr)}} &
    \multicolumn{3}{c}{\textbf{1995--2020 Recession Rate (m/yr)}} \\
    \cmidrule(lr){2-4}\cmidrule(lr){5-7}
    & \textbf{Bluff Crest} & \textbf{Bluff Toe} & \textbf{Shoreline}
    & \textbf{Bluff Crest} & \textbf{Bluff Toe} & \textbf{Shoreline} \\
    \midrule
    \textit{Ozaukee}   & -0.24 & -0.21 & -0.21 & -0.26 & -0.15 & -0.39 \\
    \textit{Milwaukee} & -0.18 & -0.05 & -0.06 & -0.22 & -0.11 & -0.26 \\
    \textit{Racine}    & -0.30 & -0.24 & -0.16 & -0.19 & -0.24 & -0.34 \\
    \textit{Kenosha}   & -0.17 & -0.28 & -0.40 & -0.14 & -0.11 & -0.24 \\
    Mean               & -0.22 & -0.17 & -0.16 & -0.22 & -0.15 & -0.32 \\
    \bottomrule
  \end{tabularx}
  \label{tab:tab2.3}
\end{table}

\begin{table}[ht]
  \caption{County-averaged beach width and bluff face slope.}
  \centering
  \small
  \setlength{\tabcolsep}{4pt}
  \renewcommand{\arraystretch}{1.2}
  \begin{tabularx}{\textwidth}{l *{4}{>{\centering\arraybackslash}p{3.3cm}}}
    \toprule
    \multicolumn{1}{c}{\textbf{County}} &
    \multicolumn{2}{c}{\textbf{Long-term average (1937--2020)}} &
    \multicolumn{2}{c}{\textbf{Short-term average (1995--2020)}} \\
    \cmidrule(lr){2-3}\cmidrule(lr){4-5}
    & \textbf{Beach width} & \textbf{Bluff face slope}
    & \textbf{Beach width} & \textbf{Bluff face slope} \\
    \midrule
    \textit{Ozaukee}   &  9.03 & 0.56 &  9.58 & 0.56 \\
    \textit{Milwaukee} & 14.66 & 0.45 & 15.12 & 0.43 \\
    \textit{Racine}    & 18.25 & 0.48 & 20.00 & 0.46 \\
    \textit{Kenosha}   &  9.51 & 0.39 &  8.02 & 0.46 \\
    Mean               & 12.98 & 0.49 & 13.59 & 0.48 \\
    \bottomrule
  \end{tabularx}
  \label{tab:tab2.4}
\end{table}

\subsection{Reach scale CGC}
Reach-level CGC is presented in Table \ref{tab:tab2.5}, while beach width and bluff face slope are summarized in Table \ref{tab:tab2.6}. All reaches exhibit consistent erosion at the bluff crest for both long-term and short-term periods. Among all reaches, reaches 7, 8, and 9—characterized by high bluffs—consistently display elevated shoreline and bluff recession rates compared to adjacent reaches. The greatest long-term recession at the bluff crest was observed in Reach 5, with a rate of –0.88 m/year from 1937 to 2020, followed by Reach 6 at –0.75 m/year from 1995 to 2020. With the exceptions of Reaches 3 and 8, most reaches experienced erosion at the bluff toe during both time periods. Reach 5 exhibited the highest toe erosion rates, at –0.82 m/year for the long-term and –0.61 m/year for the short-term. Similar to the bluff toe, shoreline positions in Reaches 3 (long-term), 6 (short-term), and 8 (long-term) showed accretion, while Reach 5 experienced highest shoreline erosion. The pronounced erosion observed at the bluff toe and shoreline in Reach 5 may be attributed to the combination of narrow beach width and steep bluff face slope (see Table \ref{tab:tab2.6}).
\begin{table}[h!]
\footnotesize
\caption{Descriptions of the 13 reaches defined for the study, reach-averaged rates of shoreline change for the 1937--2020 and 1995--2020 periods. Negative values indicate erosion, and positive values indicate lakeward movement of the shoreline. Note that Reach 1 and Reach 13 do not contain cohesive bluffs, thus no beach width is presented as no backshore feature was delineated.}
\centering
\renewcommand{\arraystretch}{1.2}
\begin{tabularx}{\textwidth}{c *{3}{>{\centering\arraybackslash}X} *{3}{>{\centering\arraybackslash}X}}
\hline
\multirow{2}{*}{Reach ID} & 
\multicolumn{3}{c}{\textbf{1937--2020 Shoreline Change Rate (m/yr)}} &
\multicolumn{3}{c}{\textbf{1995--2020 Shoreline Change Rate (m/yr)}} \\
\cline{2-7}
& Bluff Crest & Bluff Toe & Shoreline & Bluff Crest & Bluff Toe & Shoreline \\
\hline
1  & --    & --    & -0.07 & --    & --    & --    \\
2  & -0.18 & -0.14 & -0.25 & -0.34 & -0.16 & -0.67 \\
3  & -0.22 & -0.14 & -0.11 & -0.16 & -0.17 & -0.39 \\
4  & -0.32 & -0.39 & -0.38 & -0.38 & -0.15 & -0.28 \\
5  & -0.13 & -0.04 & -0.07 & -0.17 & -0.03 & -0.37 \\
6  & -0.00 &  0.15 &  0.07 & -0.06 & -0.18 & -0.30 \\
7  & -0.20 & -0.15 & -0.08 & -0.22 & -0.13 & -0.29 \\
8  & -0.69 & -0.40 & -0.36 & -0.75 &  0.13 &  0.08 \\
9  & -0.88 & -0.82 & -0.85 & -0.60 & -0.61 & -0.73 \\
10 & -0.18 & -0.09 & -0.07 & -0.07 & -0.08 & -0.21 \\
11 & -0.02 &  0.05 &  0.63 & -0.07 & -0.22 & -0.25 \\
12 & -0.15 & -0.21 & -0.29 & -0.10 & -0.12 & -0.23 \\
13 & --    & --    & -0.71 & --    & --    & -0.43 \\
\hline
\end{tabularx}
\label{tab:tab2.5}
\end{table}

\begin{table}[h!]
\caption{Reach-averaged bluff slope and beach width for the 1937--2020 and 1995--2020 periods. Negative values indicate erosion, and positive values indicate lakeward movement of the shoreline. Note that Reach 1 and Reach 13 do not contain cohesive bluffs, thus no beach width is presented as no backshore feature was delineated.}
\centering
\renewcommand{\arraystretch}{1.2}
\begin{tabularx}{\textwidth}{c *{2}{>{\centering\arraybackslash}X} *{2}{>{\centering\arraybackslash}X}}
\hline
\multirow{2}{*}{Reach ID} & 
\multicolumn{2}{c}{\textbf{Long-term average (1937--2020)}} & 
\multicolumn{2}{c}{\textbf{Short-term average (1995--2020)}} \\
\cline{2-5}
& Beach width & Bluff face slope & Beach width & Bluff face slope \\
\hline
1  & --   & --   & --    & --   \\
2  & 19.80 & 0.54 & 18.48 & 0.53 \\
3  &  6.18 & 0.52 &  7.13 & 0.51 \\
4  &  7.09 & 0.63 &  7.58 & 0.64 \\
5  & 12.69 & 0.42 & 13.09 & 0.42 \\
6  & 10.68 & 0.40 &  9.79 & 0.38 \\
7  & 18.62 & 0.52 & 20.29 & 0.51 \\
8  & 14.18 & 0.48 & 15.46 & 0.45 \\
9  &  5.13 & 0.62 &  4.91 & 0.60 \\
10 & 12.40 & 0.43 & 12.92 & 0.39 \\
11 & 65.51 & 0.38 & 76.64 & 0.37 \\
12 &  9.06 & 0.43 &  8.16 & 0.47 \\
13 & --   & --   & --    & --   \\
\hline
\end{tabularx}
\label{tab:tab2.6}
\end{table}

\subsection{CGC on transect scale}
\begin{figure}
    \centering
    \includegraphics[width=0.8\linewidth]{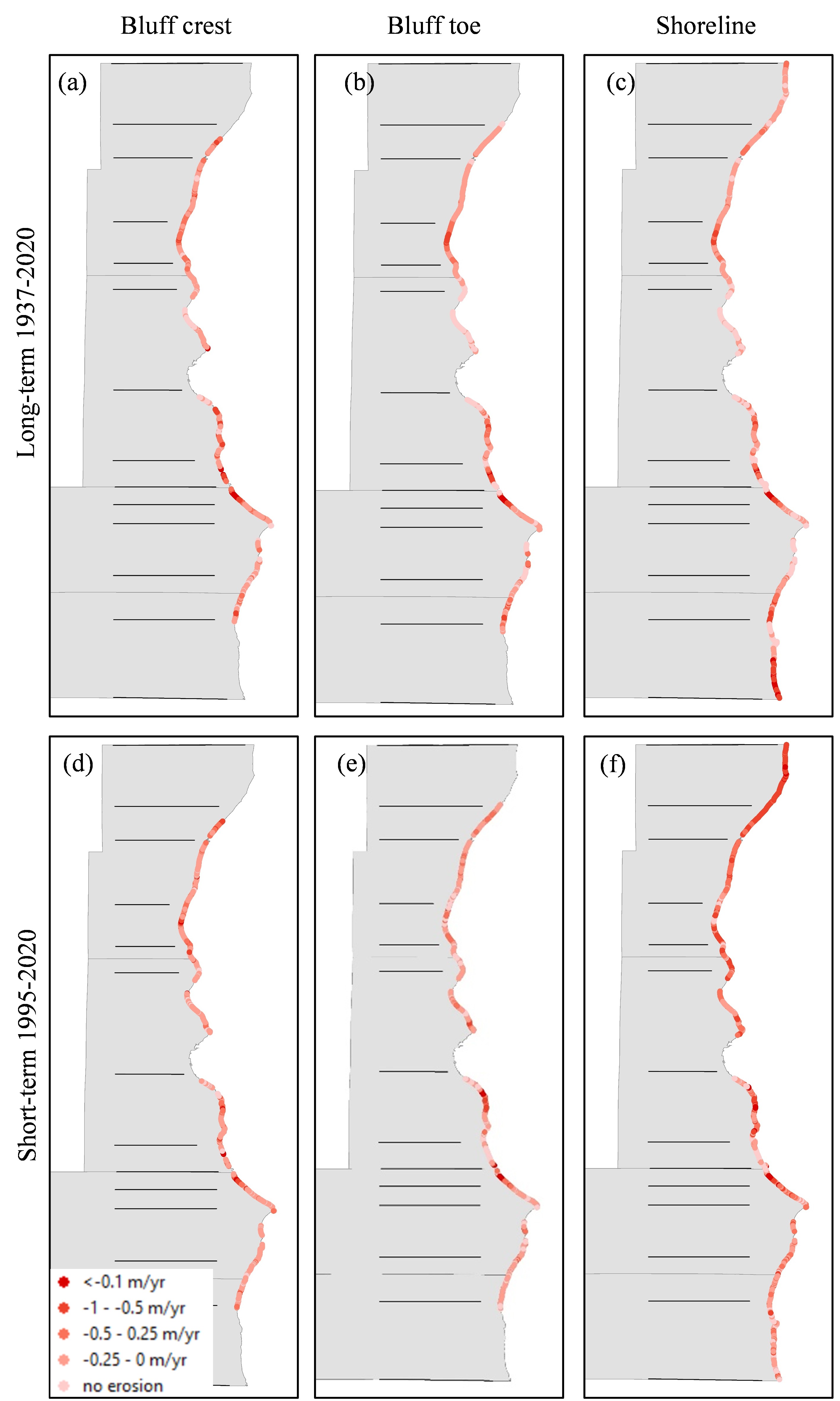}
\caption{Coastal geomorphic change in local scales. (a) long-term bluff crest,
(b) long-term bluff toe, (c) long-term shoreline, (d) short-term bluff crest,
(e) short-term bluff toe, (f) short-term shoreline.} 
\label{fig:fig2.3}
\end{figure}

Local bluff recession rates are shown as 100-meter resolution erosion maps in Fig. \ref{fig:fig2.3}. Negative values indicate landward retreat (erosion). The maximum long-term recession (1937–2020) occurred in Milwaukee County—specifically in Reach 8—with localized rates of –2.8 m/year at the bluff crest, –1.5 m/year at the toe, and –1.6 m/year at the shoreline (Fig. \ref{fig:fig2.3}a–c). Over this period, 58\% of transects recorded bluff toe retreat exceeding 0.1 m/year (i.e., more negative than –0.1 m/year), and 22\% exceeded 0.3 m/year. At the crest, 66\% of transects retreated more than 0.1 m/year and 27\% more than 0.3 m/year, while at the shoreline 58\% and 27\% of transects exceeded these thresholds, respectively. In the short-term period (1995–2020), maximum bluff toe erosion was observed in Reach 9 (Racine County) at –1.4 m/year (Fig. \ref{fig:fig2.3}e). The highest crest retreat occurred in Reach 8 (Milwaukee County) at –4.4 m/year (Fig. \ref{fig:fig2.3}d), and the greatest shoreline retreat in Reach 7 (Milwaukee County) at –2.4 m/year (Fig. \ref{fig:fig2.3}f). Overall, between 1995 and 2020, 59\% of bluff toe, 60\% of crest, and 82\% of shoreline transects experienced retreat greater than 0.1 m/year (more negative than –0.1 m/year), while 18\%, 19\%, and 53\% of the respective features exceeded 0.3 m/year.

\subsection{Error measurement and propagation for coastal geomorphic change}
Positional errors in a transect are primary from three sources: photo quality/resolution, quality of georeferencing, and digitization errors, and total error for a transect at a given year is calculated as the square root of summation of power of each error \cite{del_rio_error_2013}. For recession rate uncertainty over a period (e.g., short-term erosion rate) is given as the 95\% confidence interval on the linear regression slope \cite{thieler2009digital}. Long-term CGC rate uncertainty is 0.044 m/year for each transect in Ozaukee County, 0.028 m/year in Racine County, 0.025 m/year in Milwaukee and 0.033 m/year in Ozaukee Counties, while short-term CGC rate uncertainty is 0.034 m/year along the whole study area. The uncertainty propagation at reach and county level are summarized in Table \ref{tab:tab2.2} and Table \ref{tab:tab2.6}.

\section{Discussions}
\subsection{Role of wave impact height}

\begin{figure}
    \centering
    \includegraphics[width=0.8\linewidth]{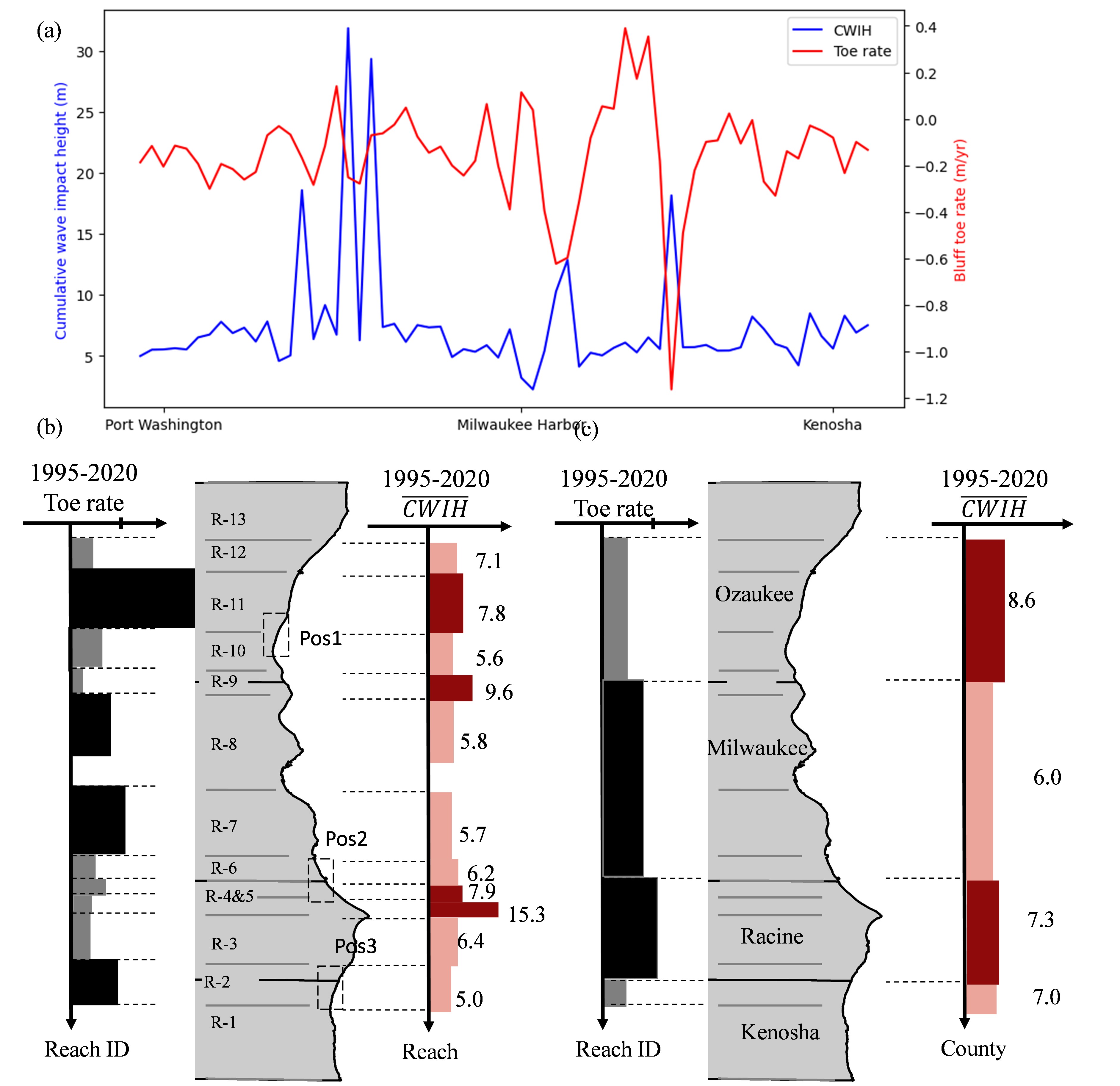}
\caption{Bluff toe recession rate and cumulative wave impact height for local
(a), reach (b) and county (c) scales for the period of 1995-2020. Heavy and mild
erosion are distinguished as values greater/less than 0.2 m/year, respectively,
and large/small (CWIH) are differentiated by values greater/less than 0.5
m/day.} 
\label{fig:fig2.7} 
\end{figure}

Wave Impact Height (WIH) is a metric from Brown \cite{brown_factors_2005} for the combined forces of water level and wave action at the bluff toe, which is presented in the following equation.
    $$WIH=SWL+ W_s+R^*-TOE$$
Where $SWL$ is the still water level (obtained from NOAA Station 9087057), $W_s$ is the wind setup, $R^*$ is wave runup in absence of bluff, and TOE is the bluff toe elevation. In this work, water levels were obtained from NOAA Guage 9087057, wind setup and wave runup were derived from Wave Information Study (WIS) hindcast data in nearby offshore cites, and toe elevations were extracted from USACE NCMP Topobathy Lidar Bathymetry. To account for the accumulative erosive impact, $\overline{CWIH}$is adapted as the sum of positive portions of hourly WIH (Swenson et al., 2006). Several other studies have found $\overline{CWIH}$ or other similar metrics of wave forcing to correlate with bluff or shoreline erosion \cite{ruggiero_wave_2001,brown_factors_2005,lin_field_2014}. Though $\overline{CWIH}$ alone cannot explain bluff behavior at an individual transect location with statistical accuracy \cite{swenson_bluff_2006}, aggregated results at broader spatial scales provide a clearer picture (Fig. \ref{fig:fig2.7}). At the reach scale, large $\overline{CWIH}$ values correspond to higher bluff toe recession rates for reaches 3 and 5, though several reaches with relatively higher wave impact demonstrated little bluff toe recession from 1995-2020 (e.g., reaches 10). However, at the county scale (Fig. \ref{fig:fig2.7}c) and local scale (Fig. \ref{fig:fig2.7}a), $\overline{CWIH}$ values do not provide a clear trend in which larger $\overline{CWIH}$ corresponds to increased toe recession. 
We hypothesize that the increasing predictive capability of $\overline{CWIH}$ when considered in the context of spatial scale is due to two main factors. First, the parameters considered in the calculation of $\overline{CWIH}$ gain greater spatial uniformity with scale. Shoreline orientation, bluff toe elevation, and nearshore slope will have some degree of local variation, but tend to fall around a central value at larger scale, while the Lake Michigan water level is assumed spatially constant over the study area at a given time. Second, the geomorphic processes governing bluff and beach erosion occur at regional and reach scales, are impacted by processes in neighboring areas and geologic inheritance, and may be subaqueous or subaerial in nature. Thus, it must be recognized that $\overline{CWIH}$, while useful in understanding hydrodynamic forcing over time, is scale-dependent in its ability to provide insight into coastal geomorphic behavior. Further, $\overline{CWIH}$ does not explicitly account for human influences, geologic differences, or sediment budget processes that are vital to understanding the geomorphic evolution of coastlines at the local level. 
\subsection{Hotspots of coastal change and causality}
\begin{figure}
    \centering
    \includegraphics[width=0.8\linewidth]{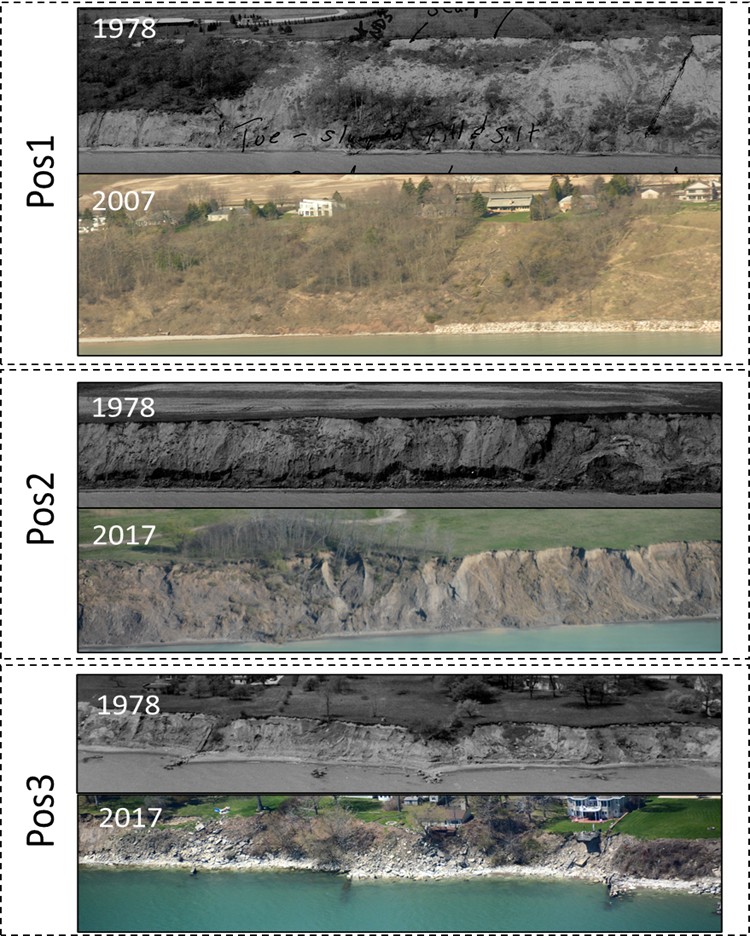}
\caption{Bluff toe recession rate and cumulative wave impact height for local
(a), reach (b) and county (c) scales for the period of 1995-2020. Heavy and mild
erosion are distinguished as values greater/less than 0.2 m/year, respectively,
and large/small (CWIH) are differentiated by values greater/less than 0.5
m/day.} 
\label{fig:fig2.8} 
\end{figure}

Fig. \ref{fig:fig2.8} shows locations of three “hotspots” with long-term bluff toe recession rates > -0.6 m/year, each of which illustrates distinct processes driving CGC in southeastern Wisconsin. Position 1 (Pos. 1 in Fig. \ref{fig:fig2.8}) is located in Reach 10 in Ozaukee County, with few coastal defense structures. The bluff toe in this location eroded at a rate of -0.63 m/year from 1937-2020, compared to a reach-average rate of -0.09 m/year. The undefended reach that contains Pos. 1 stabilized somewhat since 1995, though erosion continued from 2000-2005 at rates of -0.63 and -0.41 m/year at the crest and toe, respectively (Lin and Wu, 2014). To protect infrastructure at the bluff top, a 900 m-long revetment was constructed in 2007 to arrest bluff recession, preventing further change to the bluff face but initiating erosion along the nearshore profile in the form of lakebed downcutting. Additionally, rapid erosion took place immediately south (down-drift) of the structure, initiating further armoring and subsequent flanking erosion. Our measurements show severe erosion at the downdrift end of the terminal revetment between 2010 and 2015, with a bluff toe erosion rate of -0.91 m/year and shoreline erosion rate of -3.13 m/year.
Position 2 (pos. 2) occupies an area in Reach 5 in Racine County with no coastal defense structures, in part due to the northwest/southeast orientation of the local coastline. Structures at a large power plant protrude into the lake immediately north (up-drift) of the area, effectively cutting off littoral sediment supply to the reach. Further, the reach upstream (Reach 6) has been extensively defended, and several accretionary beaches have formed up-drift of structures. Consequently, beaches at pos. 2 within the reach are extremely narrow, and bluffs within the area have eroded at rates of -1.15 and -1.08 m/year at the crest and toe, respectively, from 1937-2020. Erosion between 1995 and 2020, a period of low water levels bookended by elevated water levels, was -0.75 and -0.57 m/year at the crest and toe, respectively. These rates are among the highest in the study area for any time period. Nevertheless, the lack of beach sediment in the area suggests that updrift near-field and far-field shoreline armoring contributes to supply-limited littoral transport, beach erosion, and accelerated bluff erosion. 
Position 3 (Pos. 3), located in Reach 2 at the Racine/Kenosha County line, illustrates a third case in which rapid bluff erosion occurred in a highly developed reach. Despite a longstanding groin field in the area, beaches remain relatively narrow, indicating low sediment supply or ineffective sediment trapping. Wave impact is comparatively high within the area and may be caused by nearshore steepening and continued beach loss. Within the groin field, bluff erosion continued at rates of -0.86 and -0.83 m/year at the bluff crest and toe, respectively, from 1937-2020. Bluff recession abated somewhat in the period of 1995-2020, with an average toe retreat rate of -0.18 m/year. Following the widespread coastal construction within southeastern Wisconsin, a pattern has emerged in which undefended portions of the coast rapidly erode, while defended areas remain stable, creating highly variable erosive conditions. For example, although the 1995-2020 average bluff toe recession rate in Reach 2 is relatively low (-0.14 m/year), 9.5\%, or approximately 1200 m, of the reach experienced bluff toe recession in excess of -0.3 m/year, spread across 16 separate locations. This highly fragmented trend is apparent in data along other highly defended portions of the coast with low average erosion rates or accretional average rates, including Reaches 1, 3, 4, 6, and 7.

\subsection{Multi-factor clustering of CGC on different spatial scales}
\begin{figure}
    \centering
    \includegraphics[width=0.8\linewidth]{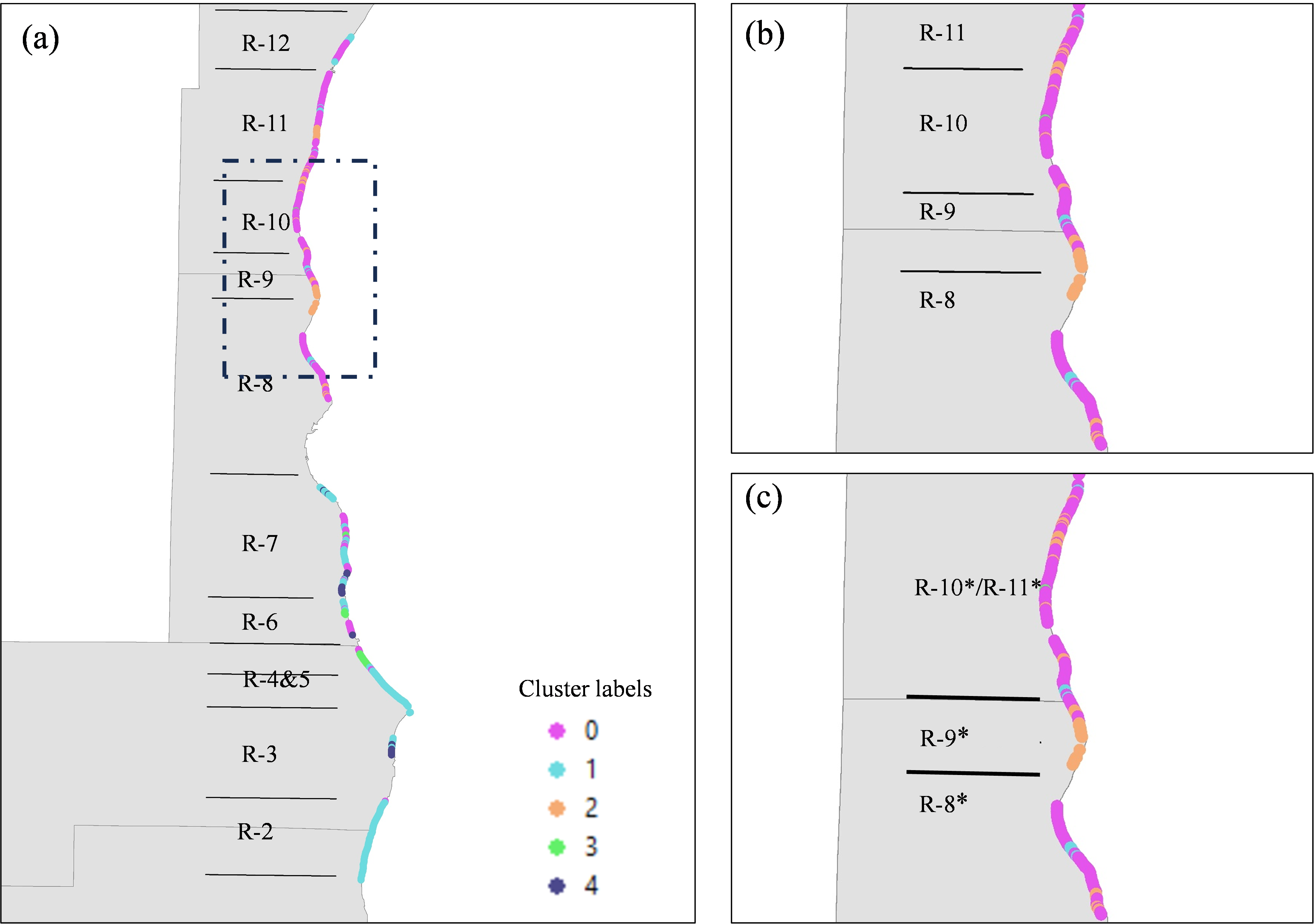}
    \caption{Multifactor clustering analysis. (a) Clustering of coastal geomorphological features; (b) zoom view of Reach 8,9,10, and 11; (c) proposed new division of reach 8*,9*,10*/11*.}
    \label{fig:figA2}
\end{figure}

Key geomorphological characteristics of CGC include shoreline and bluff change rates, bluff face slope, and beach width \cite{swenson_bluff_2006}. To better manage these characteristics for a shared strategy, reaches are divided and grouped to classify regions exhibiting similar geomorphological features \cite{shipman2008geomorphic}. The current reach delineations were originally proposed by Mikelson \cite{mickelson1977shoreline}, based on underlying geological settings. To evaluate the spatial uniformity of coastal geomorphic characteristics, a clustering analysis was conducted using long-term and short-term change rates of the bluff toe, bluff crest, shoreline, and beach width, along with bluff height, measured across 100-meter transects. To reduce dimensionality, Principal Component Analysis (PCA) was performed, and the top three principal components—accounting for about 70\% of the total variance—were used as input for the K-means clustering algorithm. The number of clusters (K) was tested across a range from 3 to 14, with five clusters identified as optimal based on clustering performance metrics. Fig. \ref{fig:figA2}a presents the results of clustering after PCA reductions from northern Kenosha County (Reach 2) to northern Ozaukee (Reach 12). Reaches 1 and 13 were omitted in the analysis due to the absence of bluffs in these areas. At the county scale, Kenosha County exhibited the highest clustering uniformity, with 100\% of transects assigned to a single cluster. However, Kenosha accounted for only 8\% of the total transects due to the exclusion of non-bluff areas. In contrast, Milwaukee County showed the lowest clustering uniformity, with the largest cluster comprising only 33\% of transects. At the reach scale, Reach 10 demonstrated the highest uniformity, with 97\% of transects assigned to the same cluster, while Reach 8 exhibited the lowest, with only 46\% of transects falling within a single cluster. Clustering results demonstrate greater uniformity at the reach scale (average of 79\%) compared to the county scale (average of 71\%). This higher consistency at the reach level highlights the importance of conducting multiscale analyses for effective coastal management. County boundaries are typically defined based on historical or political considerations and may not reflect coastal geomorphological characteristics. 

Conversely, the reach-level division is a deliberately designed coastal zone based on similar coastal characteristics, making it a more appropriate and adaptable framework for effective coastal management. For instance, this approach is particularly advantageous when assessing coastal hazard risks, as the relative homogeneity of local conditions within each reach allows for reasonable assumptions of similar risk levels, thereby reducing the complexity and effort required for evaluation \cite{torresan2008assessing,tsaimou2023impact}. Nevertheless, an inappropriate reach design may result in misclassification or grouping for coastal features, potentially misleading engineers and analysts during subsequent evaluations. For instance, as illustrated in Fig. \ref{fig:figA2}b, Reaches 10 and 11 exhibit comparable clustering characteristics and could be more effectively represented as a single unified reach. In contrast, Reaches 8 and 9 demonstrate pronounced spatial heterogeneity between their northern and southern sections, indicating the necessity for further refinement or subdivision. To improve the division of reaches, Fig. \ref{fig:figA2}c presents a revised division in which Reaches 10 and 11 are merged to form a single reach with more consistent coastal characteristics, while the boundaries of Reaches 8 and 9 are adjusted to achieve greater uniformity in coastal features. In summary, a multifactor clustering framework offers a systematic and data-driven method for reach classification, thereby enhancing the accuracy and strategic value of coastal management practices.
\subsection{Application in management, limitations and future work}
This study provides public with a dataset of coastal geomorphic changes. Part of this dataset (up to 2015) was published into the Wisconsin Shoreline Inventory \& Oblique Photo Viewer . Beyond serving as a scientific archive, the Viewer has become a practical management tool: it is incorporated into the Wisconsin State Hazard Mitigation Plan to support flood risk assessments, mitigation project scoping, and FEMA grant applications. At the county and municipal scale, the Viewer is used to communicate erosion hazards to property owners and is cited in local hazard mitigation planning documents for Milwaukee and Racine Counties, where it provides baseline evidence for prioritizing vulnerable reaches and informing shoreline stabilization strategies. By integrating long-term imagery, inventories, and quantitative recession rates, the Viewer bridges scientific datasets with coastal management practice, enabling evidence-based decision making across multiple levels of governance.

This study has two primary limitations. First, the analysis of CGC is based predominantly on historical aerial imagery, which is constrained by both the availability and quality of the data. Only seven temporal snapshots—spanning the 1930s, 1950s, 1970s, 1990s, 2000s, 2010s, and 2020s—were included; imagery from the 1940s, 1960s, 1980s, and post-2020 was unavailable. In particular, the absence of recent imagery after 2020, due to unavailability of SWERPC imagery, limits current research and engineering applications, highlighting the need for future efforts focused on data acquisition (e.g., extend imagery after 2020), characterization (e.g., automatic coastal characterization by AI), and analysis (e.g., incorporate water level and wave climate). Additionally, the temporal resolution of the dataset is uneven, with imagery acquired approximately every five years after the 1990s, but only at decadal intervals prior to that period. The unavailability and uneven distribution of data may constrain analyses, for instance, hindering efforts to reconstruct historical shoreline changes or to evaluate climate-induced erosion processes during those intervals. Second, the influence of water level fluctuations on shoreline change rates could not be fully accounted for. Variations in lake levels can influence how shoreline positions are interpreted from historical aerial imagery. In southeastern Lake Michigan, water levels rose by about 2 m between 2013 and 2020 (Gronewold et al., 2021). With an average beach slope of 0.072, this increase could exaggerate shoreline retreat by approximately 5 m/yr within this period (Troy et al., 2021). To address this limitation, future analyses should incorporate bathymetric profiles corresponding to the same years as the imagery (which was not available in most of years) and calibrate shoreline positions relative to a standardized water level. This approach would allow shoreline erosion rates to more accurately reflect changes in coastal morphology and be less affected by variations in water level.

\subsection{Conclusions}
In this study, we present the results of an inventory of coastal geomorphic change (CGC) along southeastern Wisconsin’s Lake Michigan coast. Positions of the bluff crest, bluff toe and shoreline are digitized from nine sets of aerial photos dating between 1937 and 2020 covering the entirety of Kenosha, Racine, Milwaukee, and Ozaukee Counties. Rates of coastal change are measured at 10 m intervals for long-term (1937-2020) and short-term (1995-2020) time periods, each of which featured sustained periods of above-average and below-average water levels in Lake Michigan. Rate information is aggregated to county, reach, and 100-m transect spatial scales. CGC results are presented at the county-averaged, reach-averaged, and transects (100 m) spatial scales for the long-term and short-term timescales. An analysis of long-term and short-term bluff recession at county and reach scales indicates a general stabilization of the coastal bluff toe and crest. In contrast, the shoreline exhibits a persistent trend of increasing erosion, likely driven by rising water levels, which underscores an emerging hazard to coastal stability. In the discussion, several investigations are carried out. First cumulative wave impact height is found to correlate positively with bluff toe recession at broader county and reach scales but does not correspond to higher bluff toe recession rate at the 100 m local spatial scale. A hotspot analysis at three locations is performed to reveal the local CGC caused by human activity. Last but not least, a multifactor clustering is conducted to evaluate the uniformity of coastal features in county- and reach-scales
This study offers valuable insights into practical coastal management through a multiscale analytical framework. First, the developed Wisconsin Shoreline Inventory and Oblique Viewer database serves as a publicly accessible resource to support adaptive planning and decision-making at both state and municipal levels. Second, the identification of erosion hotspots (e.g., Reaches 7, 8, and 9) provides critical guidance for prioritizing future interventions and mitigation efforts. Third, the application of a multifactor analysis presents an effective approach for refining reach delineation, thereby enhancing the precision and reliability of coastal management strategies.

\bibliographystyle{unsrt}  
\bibliography{ref}

\end{document}